# Predicting Anomalous Quantum Confinement Effect in van der Waals Materials

Kamal Choudhary[1,2], Francesca Tavazza[1]

1. Materials Science and Engineering Division, National Institute of Standards and Technology, Gaithersburg, MD 20899, U.S.A.
2. Theiss Research, La Jolla, CA, USA.

## ABSTRACT

Materials with van der Waals-bonding are known to exhibit quantum confinement effect, in which the electronic bandgap of the three-dimensional (3D) realization of a material is lower than that of its two-dimensional (2D) counterpart. However, the possibility of an anomalous quantum confinement effect (AQCE) exists, where the bandgap trend is reversed. In this work, we computationally identify materials for which such AQCE occurs. Using density functional theory (DFT), we compute ~1000 OptB88vdW (semi-local functional), ~50 HSE06 and ~50 PBE0 (hybrid functional) bandgaps for bulk and their corresponding monolayers in the JARVIS-DFT database. OptB88vdW identifies 65 AQCE materials, but the hybrid functionals only confirm such finding in 14 cases. Some of the AQCE systems identified through HSE06 and PBE0 are: hydroxides or oxide hydroxide compounds ($AlOH_2$, $Mg(OH)_2$, $Mg_2H_2O_3$, $Ni(OH)_2$, $SrH_2O_3$) as well as Sb-halogen-chalcogenide compounds (SbSBr, SbSeI) and alkali-chalcogenides (RbLiS and RbLiSe). A detailed electronic structure analysis, based on band-structure and projected density of states, shows AQCE is often characterized by lowering of the conduction band in the monolayer and corresponding changes in the $p_z$ electronic orbital contribution, with $z$ being the non-periodic direction in the 2D case. We believe our computational results would spur the effort to validate the results experimentally and will have impact on bandgap engineering applications based on low-dimensional materials.

Corresponding author: kamal.choudhary@nist.gov

## I. INTRODUCTION

Quantum confinement effect (QCE) is a well-known physical phenomenon where the electronic bandgap increases as the dimensionality of a system decreases, such as in quantum-dots[1, 2], nanotubes[3] and two-dimensional (2D) materials[4-6]. QCE occurs in low-dimensional materials because the characteristic wavelength of a material in a crystal-direction is comparable to the de Broglie wavelength of electrons[7]. The search for anomalous quantum confinement effect[8] has recently gained interest because of optoelectronic applications[9], bandgap engineering, and designing of 2D-heterostructures[10] with suitable properties[11]. Searching for such systems experimentally is unfeasible because of the large material combinatorics and high cost of each experimentation. Hence, computational methods, such as density functional theory and hybrid approaches, are the only viable option. While recently several large-scale electronic property databases have been developed[12], a systematic investigation geared towards finding AQCE has not been carried out yet, to our knowledge. Additionally, most of these 2D-material[13-15] databases are based only on semi-local density functional theory (DFT), which is known to give the right trends but incorrect bandgap description of the electronic structure[16]. Therefore, to get results better comparable to experiments, is necessary to use high level DFT methods, especially for 2D materials. There have been a few $G_0W_0$ based systematic investigations for 2D-materials, such as that by Rasmussen et al.[17] but they lack bandgaps of corresponding 3D counterparts, which are essential for establishing AQCE effect. Therefore, in order to facilitate the search for AQCE materials, high level consistent DFT calculations of both bulk and monolayer are needed.

In this work, we use the electronic structure information of both bulk and monolayers contained in the JARVIS-DFT database as starting point. The JARVIS-DFT is developed at National Institute of Standards and Technology (NIST) as a part of the Materials Genome Initiative (MGI) base NIST-JARVIS infrastructure[18], and consists of DFT evaluations of exfoliability[19], elastic[20], optoelectronic[21], topological[22], solar-cell efficiency[23], and thermoelectric[24] properties of ordered solid materials. Starting from their bulk-counterparts, monolayers are generated using the lattice-constant error and data-mining approach[19, 20]. We first compare semi-local OptB88vdW[25]-based bandgaps for bulk and monolayers, to find initial signatures of AQCE. This step is critically important because carrying out HSE06-type DFT calculations for all possible 2D materials is very time consuming. We focus on materials with low exfoliation energies only, in order to have systems that can be experimentally made into 2D form. Then, we carry out HSE06 and PBE0-type calculations on both bulk and monolayer realizations of materials previously identified as potential AQCE. We carry out these calculations on similar exfoliable compounds as well (other hydroxides or oxide hydroxides compounds, for instance). To date, we have carried out consistent OptB88vdW bandgap calculations for ~1000 materials while we have PBE0 and HSE06 data for ~50 compounds, with the database still expanding. This systematic data set of PBE0 and HSE06 electronic structure information is made publicly available on the JARVIS-DFT website (https://jarvis.nist.gov/jarvisdft), to provide an useful resource with beyond-conventional DFT results.

In addition to computing band gaps, we compare bulk and monolayer band structures and projected density of states, to investigate the mechanism behind AQCE behavior. Experimental synthesis and measurements are needed to support our computational finding, but this work acts as a first step in finding and understanding AQCE materials.

## II. METHODS

The DFT calculations were carried out using the Vienna Ab-initio simulation package (VASP)[26, 27]. The entire study was managed, monitored, and analyzed using the modular workflow, which we have made available on our github page (https://github.com/usnistgov/jarvis)[28]. We use the projector augmented wave method[29, 30] and OptB88vdW functional[25], which gives accurate lattice parameters for both Van der Waals-bonded materials (vdW) and non-vdW (3D-bulk) solids[31]. Both the internal atomic positions and the lattice constants are allowed to relax until the maximal residual Hellmann–Feynman forces on atoms are smaller than 0.001 eV $Å^{-1}$. A vacuum layer of at least 20 Å is used in monolayer (ML) calculations. The k-point mesh and plane-wave cut-off were converged for each material using the automated procedure in the JARVIS-DFT[32]. To improve the band gap calculations, we utilized two hybrid functionals: PBE0 and HSE06. In PBE0[33, 34] the exchange energy is given by a 3:1 ratio mix of the PBE and Hartree–Fock exchange energies, respectively, while the correlation is completely given by the PBE correlation energy. In HSE (Heyd–Scuseria–Ernzerhof)[35], the exchange is given by a screened Coulomb potential, to improve computational efficiency. An adjustable parameter (ω) controls how short range the interaction is. HSE06 is characterized by ω=0.2, while for a choice of ω=0, HSE transforms into PBE0.

The 2D materials are generated isolating a single monolayer, as detailed in our previous work[20]. At the beginning of this investigation, the JARVIS-DFT database contained 733 materials with exfoliation energy less than 200 meV/atom. Starting from those, we compare the bulk and the monolayer (ML) bandgaps, identifying 65 cases of possible AQCE. For some systems, the bandgaps differences are small, on the order of 0.5 eV or less, while for others they are on the order of 1 eV to 2 eV. This step is critically important because carrying out HSE06-type calculations for all possible 2D materials would be too costly in terms of computational time. We choose to consider as bandgap the minimum between the one determined using the high-density k-point mesh and the one determined investigation the high symmetry directions in the Brillouin zone. This choice was dictated by the fact that, while in 3D the Kronig-Penney model suggests that the gap should occur along high symmetry lines, in lower dimensions it can be found outside those, as in the case of 1D chain with nearest-neighbor hopping terms and next-nearest-neighbor anti-hopping terms. We also investigated exfoliable compounds of similar chemistry/composition to those identified as AQCE using OPTB88vdW (other hydroxides or oxide hydroxides systems, for instance).

For both bulk and monolayer, we run the HSE06 and PBE0-type calculations to determine the self-consistent charge density and wavefunction, using the OptB88vdW relaxed structure and k-points mesh. Then, using the generated wavefunction and charge distribution, we carry out the band structure calculation in a non-self-consistent way. We use Gaussian smearing parameter with 0.20 eV broadening for all the runs. The projected density of states is calculated using conventional Wigner size radius information.

## III. RESULTS AND DISCUSSION

The bandgap is one of the most important quantities in investigating optoelectronic properties of materials. In this work we use three methods to estimate bandgaps: OptB88vdW, HSE06 and PBE0, and apply them to both bulk (3D) and monolayer (2D) calculations. As a starting point, we analyze the OptB88vdW band gap distribution for all the exfoliable materials currently available in the JARVIS-DFT database. Currently, we have 1105 2D monolayers in the database, among which we have calculated exfoliation energies for 819 materials. Out of these 819 materials, 733 have exfoliation energies < 200 meV/atom indicating they should be easily exfoliable. Out of these 733 exfoliable 2D materials, 530 are metallic while 203 are non-metallic. In addition to hosting 2D monolayers, there are ~40000 3D materials in the JARVIS-DFT as well. As mentioned above, there are fewer non-metallic/semiconducting systems than metallic ones. In our search for AQCE materials, we focus on non-metallic systems. A flow-chart showing the AQCE identification process is shown in Fig. 1.

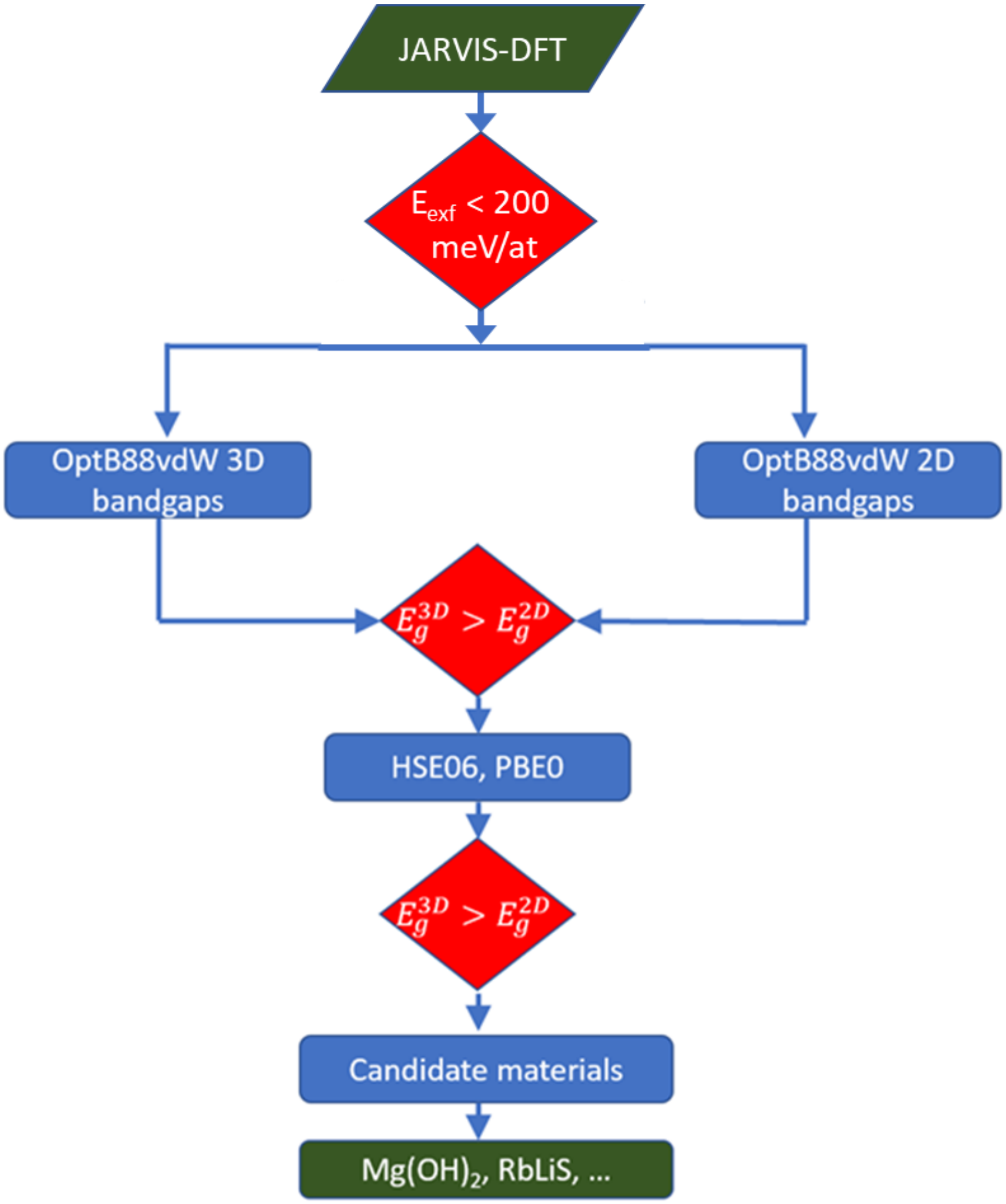

*Figure 1: Workflow leading to the identification of possible AQCE materials*

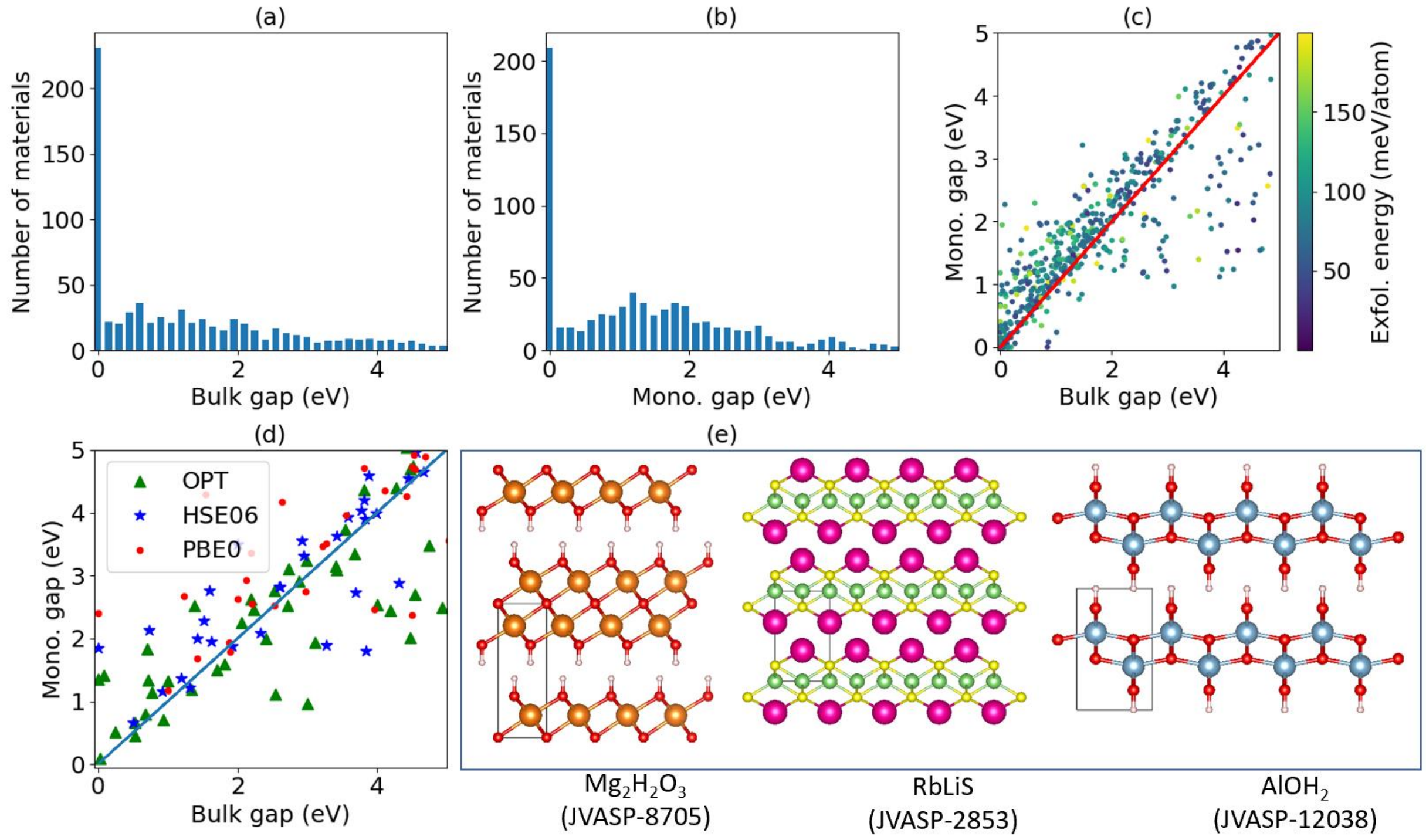

*Figure 2 Bandgap distribution of exfoliable materials in a) bulk form, b) monolayer form, c) comparison of bulk vs monolayer OptB88vdW gaps with color coded exfoliation energies, d) bulk vs monolayer bandgaps using three different functionals, e) examples of structures for materials found to display AQCE.*

We show the bandgap distribution for vdW-bonded materials in their 3D bulk and 2D monolayer configurations in Fig.2 a-b, respectively. Importantly, Figure 2b) shows that the monolayers have gap ranging from almost zero to 5 eV and more. In Fig. 2c we compare the 3D and 2D OptB88vdW bandgaps for all exfoliable materials in JARVIS-DFT with exfoliation energy as the color bar. As expected, for most cases the 2D bandgap is larger than the 3D one (92% cases), which is shown by having most of the dots above the y=x line in the scatter plot. However, we also have cases where the dot is found below the y=x line, therefore indicating possible AQCE materials. For 49 of these materials, we performed HSE06 and PBE0 calculations for both bulk and monolayer form, and Figure 2d) shows our findings, together with the correspondent OptB88vdw values for comparison. Out of 49 materials, we found 14 cases for which all three computational methods (OptB88vdw-DFT, HSE06 and PBE0)) found anomalous quantum confinement. Their list and correspondent energy gaps are given in Table 1. It is interesting to note that most AQCE-likely materials are hydroxides or oxide hydroxides compounds ($AlOH_2$, $Mg(OH)_2$, $Mg_2H_2O_3$, $Ni(OH)_2$, $SrH_2O_3$) or Sb-halogen-chalcogenides compounds (SbSBr, SbSeI) or alkali-chalcogenides (RbLiS and RbLiSe). This is very different from the conventional chalcogenide-based 2D materials, such as $MoS_2$, which are where QCE is traditionally observed. Another interesting point is that, among the 14 AQCE materials, we find examples of 0D and 1D compounds as well, not just 2D. Figure

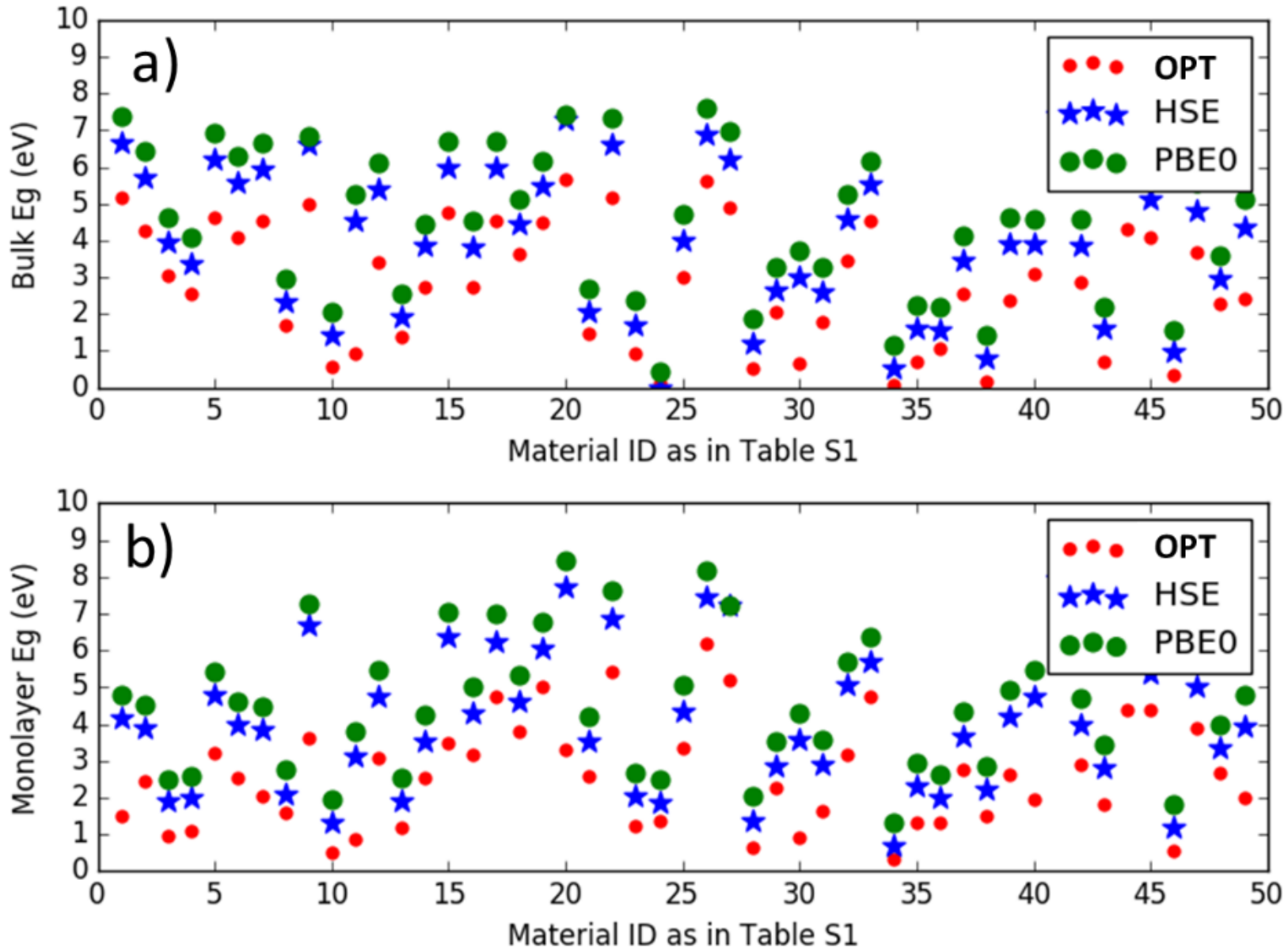

*Figure 3: Comparison of OPT, HSE06 and PBE0 band gaps. (a) 3D bulk and (b) 2D monolayer materials. The name of each material and corresponding gap values is given in Table S1, where each material is labelled using the Material ID used here.*

2e) displays some of these cases, as examples. Note that we only computed their gap in their bulk and monolayer configurations, but dimensionality criteria, like the one developed by [36], indicate some of these materials as having vdW bonding in three- (0D) or two-directions (1D), instead of just in one-direction (2D).

*Table 1 Materials that showed AQCE in all 3 theory levels. The JIDs can be used to retrieve the detailed information of a specific structure in the JARVIS-DFT database. All bandgaps are in electron-volts (eV).*

| Formula | 3D-JID | 2D-JID | 3D-gap OPT | 2D-gap OPT | 3D-gap HSE | 2D-gap HSE | 3D-gap PBE0 | 2D-gap PBE0 |
|---|---|---|---|---|---|---|---|---|
| $AlHO_2$ | JVASP-12038 | JVASP-14432 | 5.18 | 2.74 | 6.64 | 4.17 | 7.72 | 4.87 |
| $Mg_3(HO_2)_2$ | JVASP-13095 | JVASP-27956 | 4.19 | 2.45 | 5.36 | 3.55 | 6.05 | 4.18 |
| RbLiS | JVASP-2853 | JVASP-9059 | 3.01 | 1.03 | 3.84 | 1.80 | 4.50 | 2.37 |
| RbLiSe | JVASP-2928 | JVASP-9065 | 2.54 | 1.11 | 3.28 | 1.89 | 3.97 | 2.46 |
| $Mg(HO)_2$ | JVASP-4119 | JVASP-27958 | 4.55 | 2.71 | 5.92 | 4.50 | 6.49 | 5.25 |
| $Mg_2H_2O_3$ | JVASP-8705 | JVASP-27957 | 4.00 | 2.54 | 5.21 | 3.65 | 5.91 | 4.27 |
| $Mg_4H_2O_5$ | JVASP-13094 | JVASP-27955 | 4.48 | 2.45 | 5.59 | 3.51 | 6.28 | 4.12 |
| SbSBr | JVASP-5191 | JVASP-6409 | 1.74 | 1.51 | 2.33 | 2.09 | 2.98 | 2.75 |
| KCN | JVASP-14011 | JVASP-19996 | 4.94 | 2.50 | 6.56 | 6.69 | 7.31 | 7.30 |
| KAgSe | JVASP-3339 | JVASP-8867 | 0.53 | 0.46 | 1.32 | 1.22 | 1.89 | 1.79 |
| $Ni(HO)_2$ | JVASP-8753 | JVASP-27754 | 0.98 | 0.71 | 4.32 | 2.89 | 5.03 | 3.56 |
| $SrH_2O_3$ | JVASP-51673 | JVASP-75258 | 3.43 | 3.10 | 5.34 | 4.75 | 6.12 | 5.47 |
| SbSeI | JVASP-5194 | JVASP-6361 | 1.34 | 1.20 | 1.93 | 1.88 | 2.53 | 2.52 |
| SN | JVASP-4298 | JVASP-6604 | 2.73 | 2.52 | 3.70 | 3.54 | 4.43 | 4.26 |

Figure 3 shows how the energy gaps compare among the three computational approaches. As expected, OptB88vdW results are systematically underestimated with respect to the hybrid methods. PBE0 is found to be systematically higher than HSE06, as also seen in Garza and Scuseria[22]. Using both hybrid approaches is important because HSE06 reproduces experimental band gaps well in lower/mid-range band gaps[16, 37], while PBE0 does better with larger gaps[16]. This is because the amount of nonlocal exchange needed to well describe the band gap depends on the material's dielectric function, and therefore no hybrid approach, with fixed amount of Hartree-Fock exchange, can accurately describe all band gap ranges simultaneously[38, 39]. Because of this, using both methodologies provide confidence in our AQCE identification over the whole gap range. Comparing 3D to 2D findings, we notice that the difference between HSE06 band gaps and OPT ones is about 20% larger in the case of monolayer than in bulk, as shown by the MAEs in Table S1. A similar result is found for PBE0 case as well. This means that semi-local DFT band gap results for monolayer systems are probably less reliable than those for bulk. The same conclusion can be drawn by the fact that, although some materials that show AQCE using OptB88vdW still retain the AQCE behavior using HSE06 and PBE0, many others don't. In all cases the largest difference between OPT and hybrid results was found for the 2D configuration, not the bulk one. Examples of these materials are BaBrCl, CaClF and $NaNO_2$. In addition to the MAE for OPT versus HSE06 and for OPT versus PBE0, for both 3D and 2D configurations, Table S1 gives the numerical values of all the gaps plotted in Figure 3 as well as the corresponding JARVIS IDs, so that each specific material can be easily looked up in the JARVIS-DFT database, if so desired.

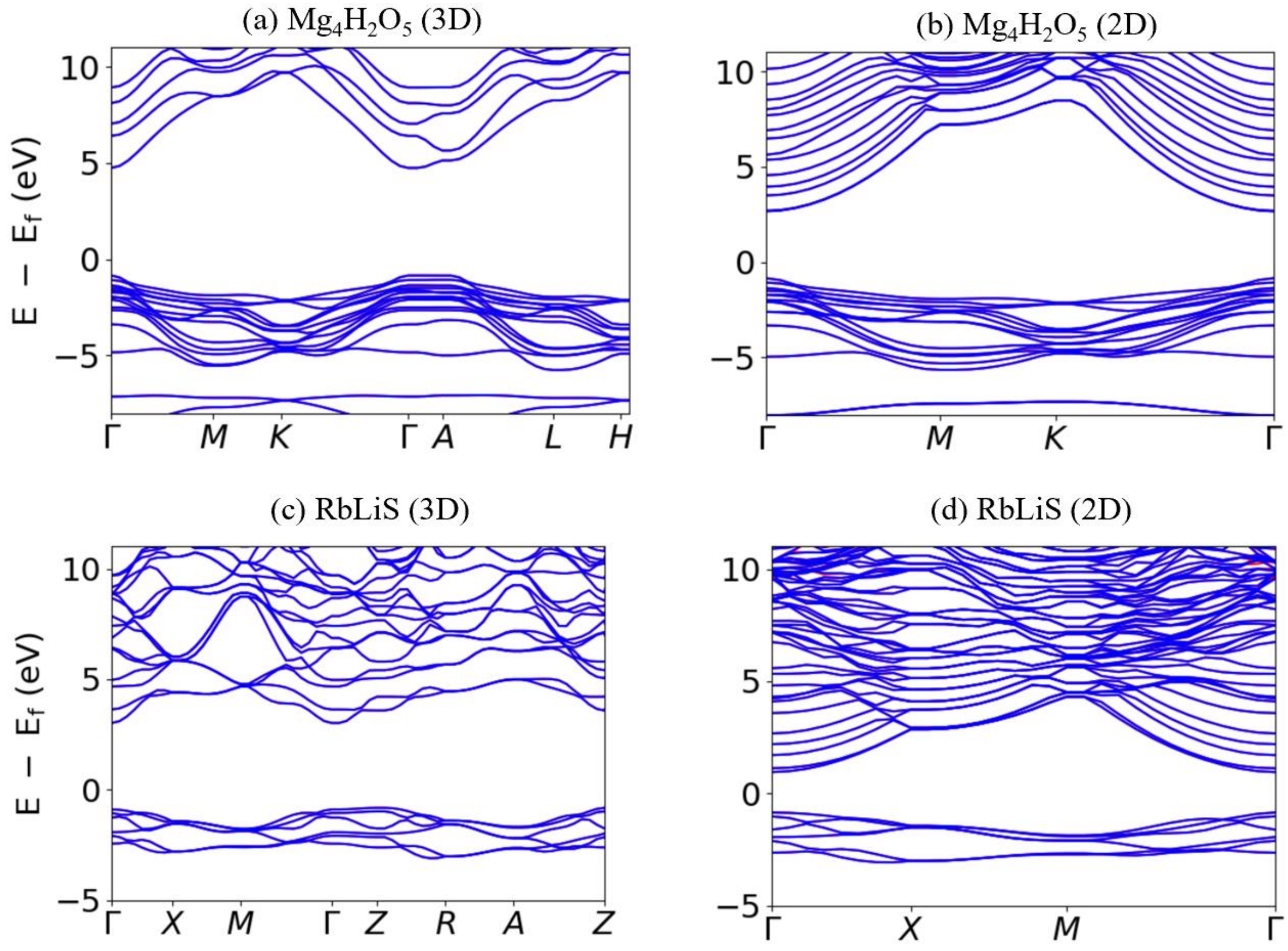

*Fig. 4 Band structure comparison of $Mg_4H_2O_5$ and RbLiS in bulk and monolayer forms.*

Next, in Figure 4 we compare the actual band structures of a couple of the systems found to display AQCE: $Mg_4H_2O_5$ as an example of the oxide hydroxides compounds and RbLiS as an example of alkali-chalcogenides. Aligning the Fermi levels, we find that for systems with AQCE the conduction band shifts down as we exfoliate a 2D material from its 3D counterpart. This is different for conventional 2D materials such as $MoS_2$ electronic structure. This suggests that unoccupied orbitals in a system can play an important role in comparing 3D vs 2D band-structures. Additionally, the bands in the 2D case seems more parabolic than the 3D case. Now, we also analyze the projected density of states of such systems to find out which orbital could be responsible for the AQCE behavior. In Fig. 5, we show such plot for the same two materials studied in Figure 4. , In the case of $Mg_4H_2O_5$ we find that the reduction in gap in the 2D monolayer is primarily due to O and H $p_z$ states as well as Mg *s* an O *s* states. Similar behavior is observed in the RbLiS case. It is important to note that the 2D materials have broken periodicity in z-direction, which justifies the difference in $p_z$ behavior between 3D and 2D counterparts. In common 2D materials the band maxima and minima are generally made of *d*-states (such as Mo-*d* states in $MoS_2$), instead of *s* or *p* ones, which may explain why in most vdW-materials the AQCE does not occur. Although existence of p-states near band extrema are a key signature, it doesn't imply that all materials with p-states contributing band extrema show AQCE. Hence, a detailed DFT calculation is needed in each case. The complete pDOS figures for the two examples under examination are shown in the SI and confirm the behavior.

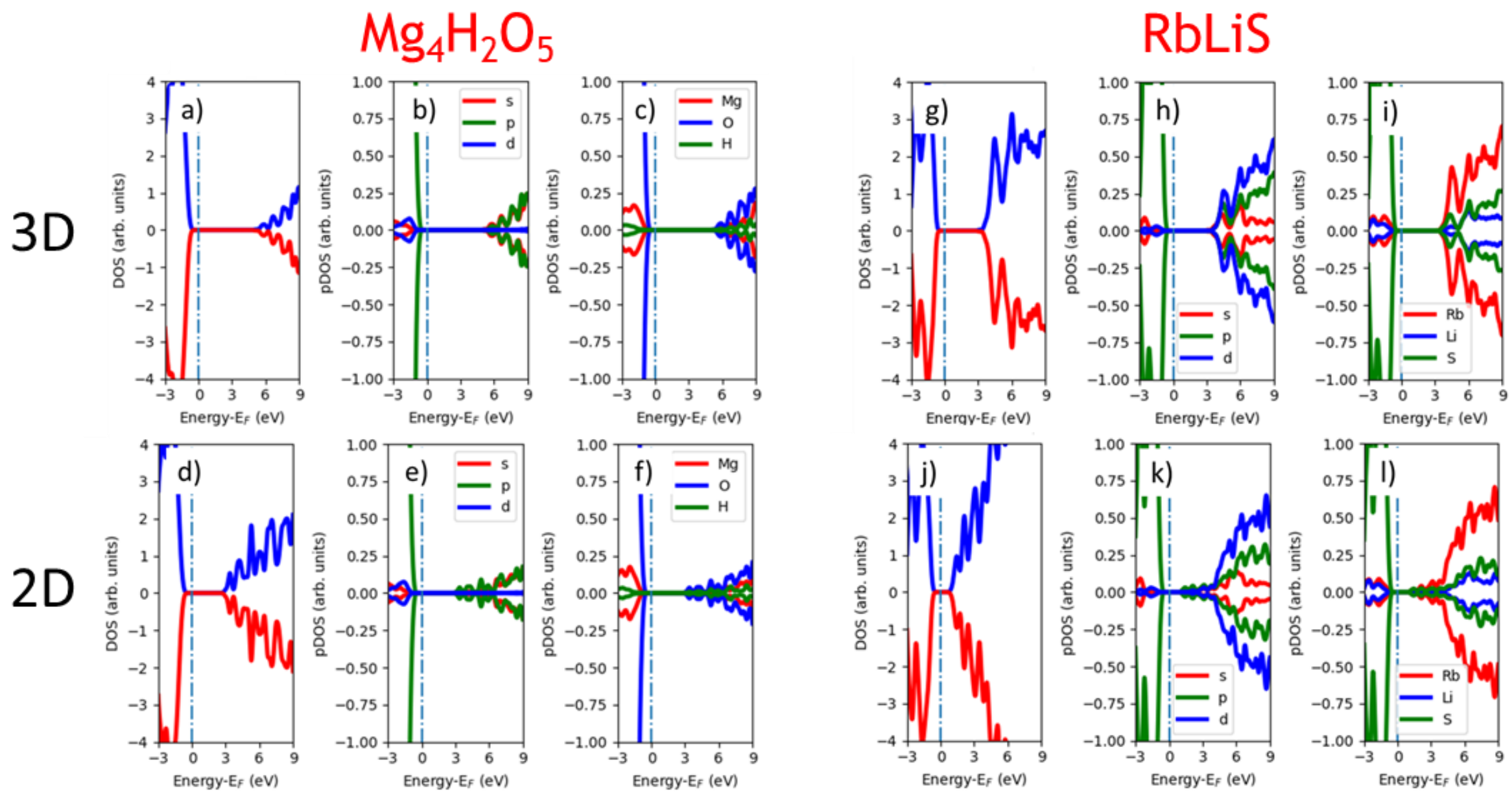

*Figure 5: HSE Band-structure comparison of bulk (3D) and monolayer (2D) for examples of AQCE systems: $Mg_4H_2O_5$ (example of hydroxide material – the most common compound type found having AQCE) and RbLiS. Total DOS (a, d, g and j), pDOS projected on orbital types (b, e, h and k) and pDOS projected on atom specie (c, f, i and l). In both cases the gap reduction in the 2D case is mostly due to p orbitals conduction band states.*

## IV. CONCLUSIONS

Using three ab-initio computational approaches (OptB88vdW, HSE06 and PBE0), we identify potential materials to host anomalous quantum confinement effect (AQCE). AQCE occurs when the three-dimensional version of a system has electronic bandgap larger than its two-dimensional (monolayer) counterpart. We first identify the signature of existence of such materials by comparing OptB88vdW based bandgaps of 3D and 2D systems available in the JARVIS-DFT database. After the initial identification, we carry out hybrid functional HSE06 and PBE0 bandgaps calculations for the candidate systems, as OptB88vdW is a semi-local method, and all such methodologies are known to underestimate bandgaps compared to experiments. Out of the ~900 systems for which JARVIS-DFT has OptB88vdW data for both bulk and monolayer structures, we compute HSE06 and PBE0 bandgaps for 49 cases, for both the bulk and the monolayer system. We find 14 materials that display AQCE in all three methodologies, and they are mostly hydroxides/oxide hydroxides compounds ($AlOH_2$, $Mg(OH)_2$, $Mg_2H_2O_3$, $Ni(OH)_2$, $SrH_2O_3$) or Sb-halogen-chalcogenides compounds (SbSBr, SbSeI) or alkali-chalcogenides (RbLiS and RbLiSe). A detailed electronic structure analysis based on band structure and projected density of states shows that the conduction band comes near to Fermi-level and $p_z$ orbital disappear in 2D-the monolayers for the AQCE systems. We believe our computational results would spur the effort to find such systems experimentally and will have impact on bandgap engineering applications based on low-dimensional materials.

## V. REFERENCES

1. Alivisatos, A. P. Semiconductor clusters, nanocrystals, and quantum dots. Science, **271**, 933-937 (1996).
2. Wise, F. W. J. Lead salt quantum dots: the limit of strong quantum confinement. Accounts Chem. Res. **33**, 773-780 (2000).
3. Chen, Z., Appenzeller, J., Knoch, J., Lin, Y.-m. & Avouris, P. The role of metal− nanotube contact in the performance of carbon nanotube field-effect transistors. Nano Lett. **5**, 1497-1502 (2005).
4. Li, S.-L. *et al.* Thickness scaling effect on interfacial barrier and electrical contact to two-dimensional MoS2 layers. ACS Nano **8**, 12836-12842 (2014).
5. Kim, T.-Y. *et al.* Quantum confinement effect of silicon nanocrystals in situ grown in silicon nitride films. App. Phys. Lett. **85**, 5355-5357 (2004).
6. Gan, Z. *et al.* Quantum confinement effects across two-dimensional planes in MoS2 quantum dots. App. Phys. Lett. **106**, 233113 (2015).
7. Mao, J., Liu, Z. & Ren, Z. J. Size effect in thermoelectric materials. npj Quant. Mat. **1**, 1-9 (2016).
8. Wang, X., Zhang, R., Lee, S. T., Frauenheim, T. & Niehaus, T. A. Anomalous size dependence of the luminescence in reconstructed silicon nanoparticles. App. Phys. Lett. **93**, 243120 (2008).
9. Habinshuti, J. *et al.* Anomalous quantum confinement of the longitudinal optical phonon mode in PbSe quantum dots. Phys. Rev. B **88**, 115313 (2013).
10. Novoselov, K., Mishchenko, A., Carvalho, A. & Neto, A. C. 2D materials and van der Waals heterostructures. Science **353** (2016).
11. Kim, D., Jung, J. H. & Ihm, J. Theoretical Study of Aluminum Hydroxide as a Hydrogen-Bonded Layered Material. Nanomat. **8**, 375 (2018).
12. Vasudevan, R. K. *et al.* Materials science in the artificial intelligence age: high-throughput library generation, machine learning, and a pathway from correlations to the underpinning physics. MRS Comm. **9**, 821-838 (2019).
13. Mounet, N. *et al.* Two-dimensional materials from high-throughput computational exfoliation of experimentally known compounds. Nature Naotech. **13**, 246-252 (2018).

14. Haastrup, S. *et al.* The Computational 2D Materials Database: high-throughput modeling and discovery of atomically thin crystals. 2D Mater. **5**, 042002 (2018).
15. Choudhary, K., Kalish, I., Beams, R. & Tavazza, F. High-throughput Identification and Characterization of Two-dimensional Materials using Density functional theory. Scientific Reports **7**, 1-16 (2017).
16. Garza, A. J. & Scuseria, G. E. Predicting band gaps with hybrid density functionals. J. Phys. Chem. Lett. **7**, 4165-4170 (2016).
17. Rasmussen, F. A. & Thygesen, K. S. Computational 2D materials database: Electronic structure of transition-metal dichalcogenides and oxides. *The Journal of Physical Chemistry C* **119**, 13169-13183 (2015).
18. Choudhary, K., Garrity, K.F., Reid, A.C.E. et al. The joint automated repository for various integrated simulations (JARVIS) for data-driven materials design. npj Comput Mater 6, 173 (2020). https://doi.org/10.1038/s41524-020-00440-1.
19. Choudhary, K., Cheon, G., Reed, E. & Tavazza, F. Elastic properties of bulk and low-dimensional materials using van der Waals density functional. Physical Review B **98**, 014107 (2018).
20. Choudhary, K. *et al.*, Density Functional Theory based Electric Field Gradient Database. Scientific Data **7**, 362 (2020).
21. Choudhary, K. *et al.* Computational screening of high-performance optoelectronic materials using OptB88vdW and TB-mBJ formalisms. Scientific Data **5**, 180082 (2018).
22. Choudhary, K., Garrity, K. F. & Tavazza, F. High-throughput Discovery of Topologically Non-trivial Materials using Spin-orbit Spillage. Scientific Reports **9**, 8534 (2019).
23. Choudhary, K. *et al.* Accelerated Discovery of Efficient Solar-cell Materials using Quantum and Machine-learning Methods. Chemistry of Materials (2019).
24. Choudhary, K., Garrity, K. & Tavazza, F. Data-driven Discovery of 3D and 2D Thermoelectric Materials. Journal Physics Condensed Matter **32**, 47 (2020).
25. Klimeš, J., Bowler, D. R. & Michaelides, A. J. Chemical accuracy for the van der Waals density functional. J. Phys. Cond. Mat. **22**, 022201 (2009).
26. Kresse, G.; Furthmüller, J., Efficient iterative schemes for ab initio total-energy calculations using a plane-wave basis set. Phys Rev. B **54** (16), 11169 (1996).
27. Kresse, G.; Furthmüller, J., Efficiency of ab-initio total energy calculations for metals and semiconductors using a plane-wave basis set. Comp.Mat. Sci. **, 6** (1), 15-50 (1996).
28. Please note that commercial software is identified to specify procedures. Such identification does not imply recommendation by the National Institute of Standards and Technology. .
29. Blöchl, P. E. , Projector augmented-wave method. **1994,** Phys. Rev. B*50* (24), 17953.
30. Kresse, G.; Joubert, D., From ultrasoft pseudopotentials to the projector augmented-wave method. **1999,** Phys. Rev. B *59* (3), 1758.
31. Choudhary, K.; Kalish, I.; Beams, R.; Tavazza, F., High-throughput Identification and Characterization of Two-dimensional Materials using Density functional theory. *Scientific Reports* **2017,** *7* (1), 5179.
32. Choudhary, K.; Tavazza, F., Convergence and machine learning predictions of Monkhorst-Pack k-points and plane-wave cut-off in high-throughput DFT calculations. Computational Materials Science *161*, 300-308 (2019).
33. Perdew, J. P.; Ernzerhof, M.; Burke, K. J. T. Rationale for mixing exact exchange with density functional approximations. J. Chem. Phys. *105* (22), 9982-9985 (1996).

34. Adamo, C.; Barone, V. J. T., Toward reliable density functional methods without adjustable parameters: The PBE0 model. J. Chem. Phys. *110* (13), 6158-6170 (1999).
35. Heyd, J.; Scuseria, G. E.; Ernzerhof, M. J. T., Hybrid functionals based on a screened Coulomb potential. J. Chem. Phys. *118* (18), 8207-8215 (2003).
36. Cheon, G.; Duerloo, K.-A. N.; Sendek, A. D.; Porter, C.; Chen, Y.; Reed, E. J., Data mining for new two-and one-dimensional weakly bonded solids and lattice-commensurate heterostructures. Nano Lett. *17* (3), 1915-1923 (2017).
37. Chen, W.; Pasquarello, A., Band-edge levels in semiconductors and insulators: Hybrid density functional theory versus many-body perturbation theory. Phys. Rev. B, *86* (3), 035134 (2012).
38. Skone, J. H.; Govoni, M.; Galli, G., Nonempirical range-separated hybrid functionals for solids and molecules. Phys. Rev. B *93* (23), 235106 (2016).
39. Marques, M. A.; Vidal, J.; Oliveira, M. J.; Reining, L.; Botti, S., Density-based mixing parameter for hybrid functionals**,** Phys. Rev B, *83* (3), 035119 (2011).

## Supplementary material: Predicting Anomalous Quantum Confinement Effect in van der Waals Materials

Kamal Choudhary[1,2], Francesca Tavazza[1]

1. Materials Science and Engineering Division, National Institute of Standards and Technology, Gaithersburg, MD 20899, U.S.A.
2. Theiss Research, La Jolla, CA, USA.

*Table S1: DFT, HSE and PBE0 band gaps for all the materials we studied, both in 3D and in 2D configuration.*

| ID in Fig.3 | Material | 3D JID | 3D Eg (eV) | | | 2D JID | 2D Eg (eV) | | |
|---|---|---|---|---|---|---|---|---|---|
| | | | OPT | HSE06 | PBE0 | | OPT | HSE06 | PBE0 |
| 1 | $AlHO_2$ | JVASP-12038 | 5.17 | 6.64 | 7.37 | JVASP-14432 | 1.50 | 4.17 | 4.81 |
| 2 | $Mg_3(HO2)_2$ | JVASP-13095 | 4.19 | 5.36 | 6.05 | JVASP-27956 | 2.45 | 3.55 | 4.18 |
| 3 | RbLiS | JVASP-2853 | 3.01 | 3.84 | 4.50 | JVASP-9059 | 0.97 | 1.80 | 2.37 |
| 4 | RbLiSe | JVASP-2928 | 2.54 | 3.28 | 3.97 | JVASP-9065 | 1.11 | 1.89 | 2.46 |
| 5 | $Mg(HO)_2$ | JVASP-4119 | 4.55 | 5.92 | 6.49 | JVASP-27958 | 2.71 | 4.50 | 5.25 |
| 6 | $Mg_4H_2O_5$ | JVASP-8705 | 4.00 | 5.21 | 5.91 | JVASP-27957 | 2.54 | 3.65 | 4.27 |
| 7 | $Mg4H_2O_5$ | JVASP-13094 | 4.48 | 5.59 | 6.28 | JVASP-27955 | 2.02 | 3.51 | 4.12 |
| 8 | SbSBr | JVASP-5191 | 1.71 | 2.33 | 2.98 | JVASP-6409 | 1.51 | 2.09 | 2.75 |
| 9 | KCN | JVASP-14011 | 4.94 | 6.56 | 6.85 | JVASP-19996 | 2.50 | 6.67 | 7.29 |
| 10 | KAgSe | JVASP-3339 | 0.53 | 1.32 | 1.89 | JVASP-8867 | 0.46 | 1.22 | 1.79 |

| | | | | | | | | | | | |
|---|---|---|---|---|---|---|---|---|---|---|---|
| 11 | $Ni(HO)_2$ | | JVASP-8753 | 0.94 | 4.32 | 5.03 | | JVASP-27754 | 0.71 | 2.89 | 3.56 |
| 12 | $SrH_2O_3$ | | JVASP-51673 | 3.43 | 5.34 | 6.12 | | JVASP-75258 | 3.10 | 4.75 | 5.47 |
| 13 | SbSeI | | JVASP-5194 | 1.34 | 1.93 | 2.53 | | JVASP-6361 | 1.19 | 1.88 | 2.52 |
| 14 | SN | | JVASP-4298 | 2.72 | 3.70 | 4.43 | | JVASP-6604 | 2.52 | 2.74 | 4.26 |
| 15 | BaBrCl | | JVASP-5443 | 4.75 | 5.97 | 6.68 | | JVASP-6748 | 3.48 | 6.08 | 7.04 |
| 16 | $SiS_2$ | | JVASP-137 | 2.73 | 3.81 | 4.53 | | JVASP-6016 | 3.11 | 4.20 | 4.93 |
| 17 | $BCl_3$ | | JVASP-164 | 4.48 | 5.92 | 6.66 | | JVASP-5905 | 4.70 | 6.16 | 6.92 |
| 18 | $MgI_2$ | | JVASP-167 | 3.55 | 4.45 | 5.07 | | JVASP-5965 | 3.74 | 4.55 | 5.17 |
| 19 | $MgBr_2$ | | JVASP-2068 | 4.43 | 5.47 | 6.16 | | JVASP-6040 | 5.04 | 5.90 | 6.57 |
| 20 | CaClF | | JVASP-22529 | 5.63 | 7.19 | 7.41 | | JVASP-60289 | 3.32 | 7.64 | 8.33 |
| 21 | PbO | | JVASP-252 | 1.38 | 2.01 | 2.64 | | JVASP-6007 | 2.52 | 3.50 | 4.18 |
| 22 | TmClO | | JVASP-30183 | 5.13 | 6.59 | 7.34 | | JVASP-60554 | 5.40 | 6.87 | 7.60 |
| 23 | $ZrS_2$ | | JVASP-82 | 0.77 | 1.62 | 2.21 | | JVASP-689 | 1.15 | 1.96 | 2.55 |
| 24 | $PtSe_2$ | | JVASP-128 | 0.01 | 0.00 | 0.00 | | JVASP-744 | 1.35 | 1.85 | 2.40 |
| 25 | $PBr_3$ | | JVASP-4279 | 2.99 | 3.99 | 4.70 | | JVASP-6091 | 3.24 | 4.00 | 4.90 |
| 26 | $MgCl_2$ | | JVASP-170 | 5.58 | 6.86 | 7.57 | | JVASP-5986 | 6.20 | 7.28 | 7.98 |
| 27 | CBrN | | JVASP-5437 | 4.89 | 6.21 | 6.94 | | JVASP-6745 | 5.21 | 7.04 | 7.05 |
| 28 | $Ta_3TeI_7$ | | JVASP-5632 | 0.53 | 1.20 | 1.88 | | JVASP-6799 | 0.66 | 1.37 | 1.94 |
| 29 | $GeI_2$ | | JVASP-179 | 2.06 | 2.61 | 3.22 | | JVASP-5998 | 2.26 | 2.83 | 3.47 |
| 30 | $MnO_2$ | | JVASP-5308 | 0.68 | 2.93 | 1.54 | | JVASP-6922 | 0.80 | 3.56 | 4.30 |
| 31 | $As_4S_3$ | | JVASP-5362 | 1.81 | 2.60 | 3.27 | | JVASP-6940 | 1.59 | 2.82 | 3.52 |
| 32 | $CaI_2$ | | JVASP-2065 | 3.41 | 4.56 | 5.26 | | JVASP-6052 | 3.16 | 4.97 | 5.62 |
| 33 | SrIF | | JVASP-3594 | 4.52 | 5.37 | 6.02 | | JVASP-13579 | 4.74 | 5.62 | 6.32 |
| 34 | $TiS_2$ | | JVASP-317 | 0.03 | 0.51 | 1.01 | | JVASP-771 | 0.10 | 0.66 | 1.17 |
| 35 | $WS_2$ | | JVASP-72 | 0.72 | 1.52 | 2.12 | | JVASP-658 | 1.34 | 2.29 | 2.93 |
| 36 | $WSe_2$ | | JVASP-75 | 1.00 | 1.42 | 2.01 | | JVASP-652 | 1.33 | 2.00 | 2.63 |
| 37 | HoSI | | JVASP-28375 | 2.53 | 3.43 | 4.12 | | JVASP-27841 | 2.76 | 3.64 | 4.36 |
| 38 | GaTe | | JVASP-4666 | 0.09 | 0.74 | 1.23 | | JVASP-6838 | 1.42 | 2.14 | 2.68 |
| 39 | $TiO_2$ | | JVASP-308 | 2.23 | 3.78 | 4.51 | | JVASP-5902 | 2.46 | 4.04 | 4.73 |
| 40 | TlF | | JVASP-2106 | 3.11 | 3.88 | 4.54 | | JVASP-6076 | 1.94 | 4.59 | 5.26 |
| 41 | $SiH_4$ | | JVASP-5281 | 6.14 | 7.40 | 8.14 | | JVASP-6544 | 6.57 | 7.87 | 8.60 |
| 42 | $CS_2$ | | JVASP-4388 | 2.88 | 3.83 | 4.54 | | JVASP-6895 | 2.92 | 3.90 | 4.70 |
| 43 | GaSe | | JVASP-81 | 0.70 | 1.60 | 2.19 | | JVASP-687 | 1.84 | 2.76 | 3.37 |
| 44 | $ZrP_2(HO_3)_2$ | | JVASP-12585 | 4.27 | 5.95 | 6.70 | | JVASP-27899 | 4.40 | 6.05 | 6.80 |
| 45 | SrHBr | | JVASP-3738 | 3.81 | 4.85 | 5.55 | | JVASP-8927 | 4.37 | 5.38 | 6.09 |
| 46 | $ZrSe_2$ | | JVASP-365 | 0.25 | 0.92 | 1.43 | | JVASP-789 | 0.51 | 1.16 | 1.69 |
| 47 | $AsCl_3$ | | JVASP-4273 | 3.68 | 4.67 | 5.29 | | JVASP-6364 | 3.35 | 4.66 | 5.69 |
| 48 | $PbI_2$ | | JVASP-161 | 2.19 | 2.95 | 3.56 | | JVASP-5995 | 2.63 | 3.32 | 3.96 |
| 49 | $NaNO_2$ | | JVASP-1429 | 2.41 | 3.58 | 3.81 | | JVASP-8987 | 2.00 | 3.93 | 4.71 |
| | Mean Abs. Dev. (MAD) | | | | 1.08 | 1.66 | | | | 1.33 | 2.01 |

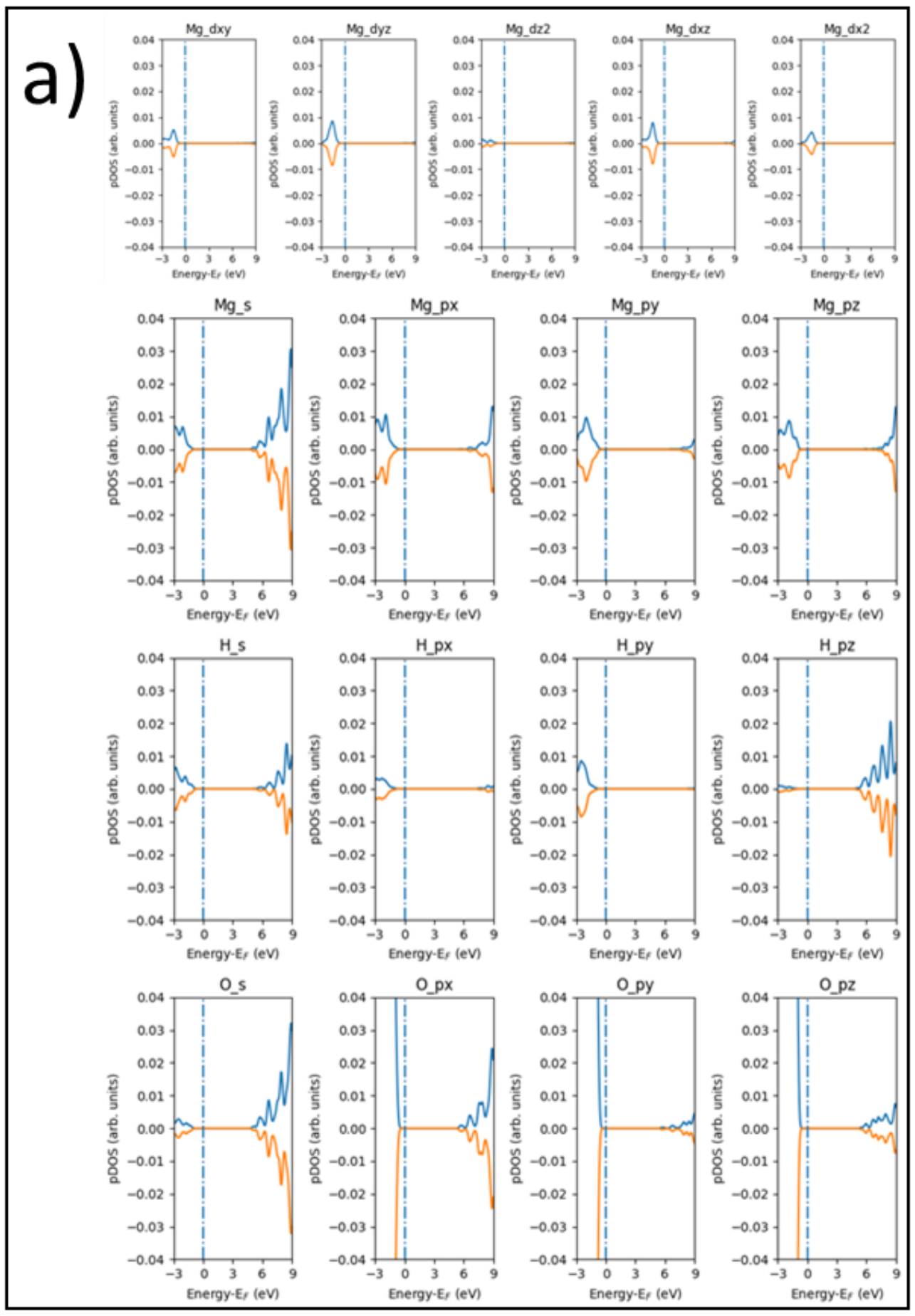

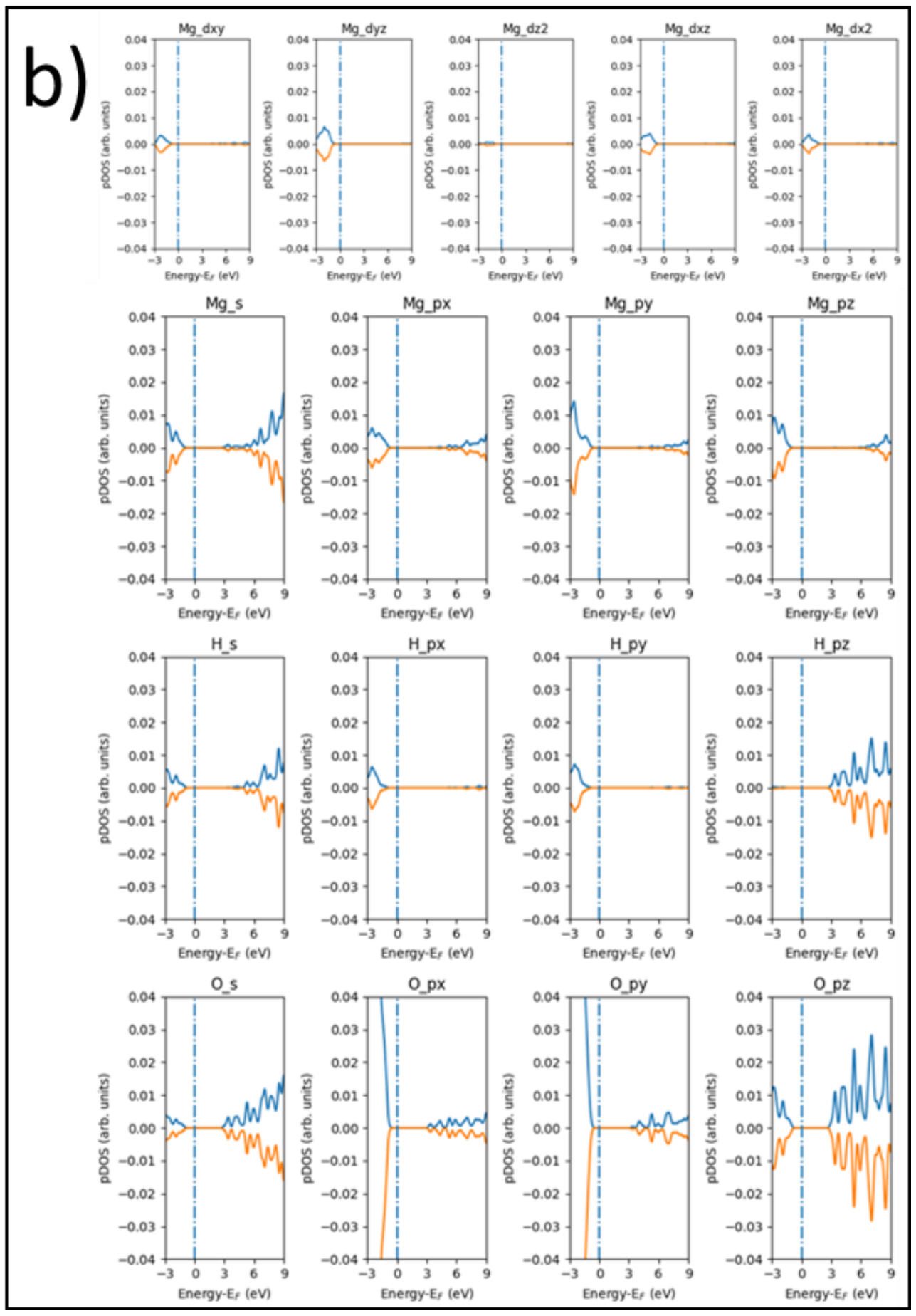

*Figure S1: $Mg_4H_2O_5$ pDOS projected on every atom-orbital combination for bulk (a) and monolayer (b) configurations.*

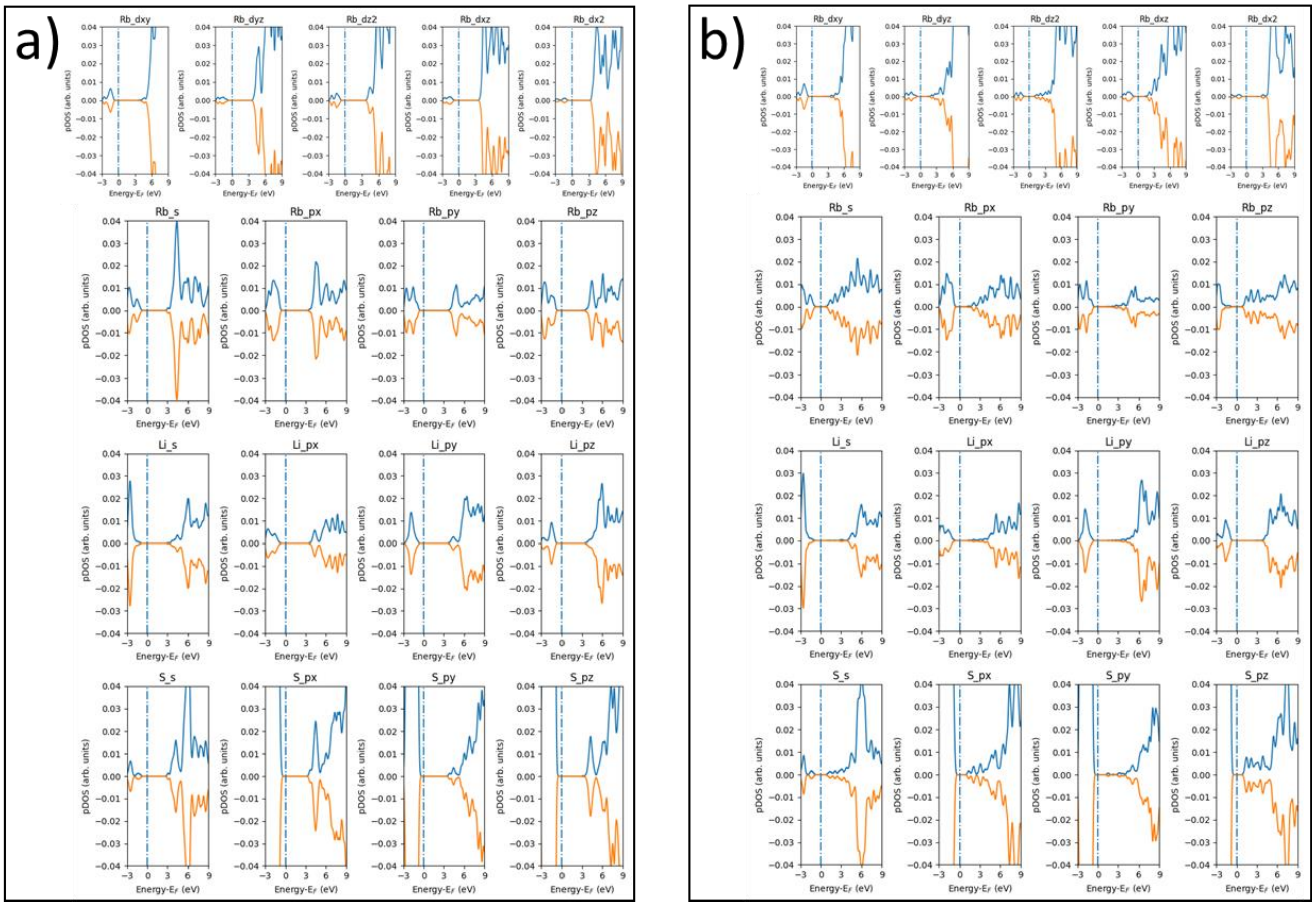

*Figure S2: RbLiS pDOS projected on every atom-orbital combination for bulk (a) and monolayer (b) configurations.*